# RELATIVISTIC GEOMETRY AND WEAK INTERACTIONS

**Gustavo González-Martín**

Departamento de Física, Universidad Simón Bolívar,

Apartado 89000, Caracas 1080-A, Venezuela.

Web page URL http:\\prof.usb.ve\ggonzalm\

Geometric interactions in a new relativistic geometric unified theory include interactions other than gravitation and electromagnetism. In a low energy limit one of these interactions leads essentially to a Fermi type theory of weak interactions.

## Introduction.

A non linear geometric theory that unifies gravitation and electromagnetism offers the possibility of representing other interactions by a sector of the connection [1]. It has been shown that new electromagnetic consequence of the theory lead to quanta of electric charge and magnetic flux, providing a plausible explanation to the fractional quantum Hall effect [2], [3]. On the other hand we also have considered an approximation to this geometric non linear theory [4] where the microscopic physical objects (geometric particles) are realized as linear geometric excitations, geometrically described in a jet bundle formalism shown to lead to the standard quantum field theory techniques. These geometric excitations are essentially perturbations around a non linear geometric background space solution, where the excitations may be considered to evolve with time. In this framework, a geometric particle is acted upon by the background connection and is never really free except in absolute empty background space (zero background curvature). The background space carries the universal inertial properties which should be consistent with the ideas of Mach [5] and Einstein [6] that assign fundamental importance of far-away matter in determining the inertial properties of local matter. We may interpret the geometric excitations as geometric particles and the background as the particle vacuum. This is a generalization of what is normally done in quantum field theory when particles are interpreted as vacuum excitations. The vacuum is replaced by a geometric symmetric curved background space and current solution which we call the substratum. The field equation of the theory admits a constant connection and current substratum solution [7], [8].

The role that Clifford algebras play in the geometrical structure of the theory provides a link to the non classical interactions theories. The structure group of the physical geometry is the simple group of automorphisms of the Clifford algebra. The combinations of the fundamental geometric excitations corresponding to the holonomy subgroups of the connection display an interesting SU(3)⊗SU(2)⊗U(1) combinatorial symmetry [9]. The standard model, that shows this symmetry, has had many successes in describing weak and strong interactions that lead to a general acceptance of the model. Nevertheless, nowadays, this model may be considered an effective theory of some other more general theory. History has shown us that in many cases, progress in physics is attained by the evolution or replacement of models, that provide partial fits to experimental data, by more general ones. The consideration of geometrical strings to represent physical concepts indicates a present trend in this direction. Therefore, in the quest of unification, the lack of an a established relation between geometry and the standard model should not deter us from investigating the possible physical interpretations of the odd sector of the connection which precisely holds the key to its relation with the standard model. This sector certainly does not represent simply the classical interactions and, in particular, may help clarify a possible relation to the standard model. As a matter of fact, the physical geometry indicates there is complementary approach to particles and their interactions related to another model [10]. As first task we consider here low energy aspects of weak interactions.

The holonomy groups of the connection may be used geometrically to classify the interactions contained in theory. The subgroup chain SL(4,R) ⊃ Sp(4,R) ⊃ SL(2,C) characterizes a chain of subinteractions with reducing sectors of non classical interactions  In this article we shall mainly concern ourselves with Sp(4,R) and SL(2,C) connections

A section is related by charts (coordinates) to elements of the group, SL(4,R), which are matrices that form a frame of general SL(4,R) column spinors. The connection is a sl(4,R) 1-form that acts naturally on the frame $e$ (sections).

The field equation, which relates the derivatives of the curvature to a current source $J$ is

$$D^*\Omega = k \, {}^*J = 4\pi\alpha \, {}^*J \, , \qquad (1)$$

$$J^\mu = \tilde{e}\,\kappa^{\hat{\alpha}} u^\mu_{\hat{\alpha}} e \, , \qquad (2)$$

in terms of the frame $e$, an orthonormal set of the algebra $\kappa$, the correlation in spinor spaces and an space-time tetrad $u$. The coupling constant is $4\pi\alpha$, where $\alpha$, is the fine structure constant. This relation introduces a fundamental length, as we shall see later.

The field equation implies a conservation law for the geometric current, which determines a generalized Dirac equation in terms of the frames. This equation, for the even an odd parts of a frame $f$ reduces, under certain restrictions [11], to

$$\kappa^\mu \partial_\mu f_+ = \kappa^\mu \Gamma_{\mu-} f_- = m f_- \, , \qquad (3)$$

$$\kappa^\mu \partial_\mu f_- = \kappa^\mu \Gamma_{\mu-} f_+ = m f_+ \ , \qquad (4)$$

implying that a frame for a massive corpuscle must have odd and even parts. In our case

$$f_- = 0 \implies m = 0 \ . \qquad (5)$$

Therefore, for an even frame we have, multiplying by $\kappa^0$,

$$\bar{\sigma}^\mu \partial_\mu f_+ = 0 \ , \qquad (6)$$

which is the equation normally associated with a neutrino field. A fluctuation of $f_+$ on the fixed background obeys also the last equation.

We also have suggested that particles may be represented by excitations on a geometric background. In particular, the electron and neutrino, at fixed states, correspond to matrices with only one non zero column which form an algebraic representation of the group.

## Geometric Weak Interaction.

If the geometric theory has anything to do with the weak interactions, it should be possible to represent electron neutrino interactions within these geometric ideas. From the discussion of previous articles, we consider that an electroweak interaction may be related to the action of the Sp(4,R) holonomy group. The total interaction field should be represented by a sp(4,R) connection $\Gamma$. The total matter current should be associated to a Sp(4,R) frame $f$ presenting both the electron field $e$ and the neutrino field $\nu$. to avoid confusion, in this article we use the symbol $f$ for a general section reserving $e$ for electron sections

At a point, the total frame $f$ of the interacting $e$, $\nu$ is related to an element of the group Sp(4,R), a subspace of the geometric algebra $R_{3,1}$,=R(4). The frame $f$ may be decomposed into fields associated to the particles $e$, $\nu$ by means of the addition operation within the algebra. These $e$, $\nu$ fields are not necessarily frames because addition does not preserve the group subspace Sp(4,R) and geometrically, $e$ $\nu$ are sections of a fiber bundle with space-time as base and the Clifford algebra as fiber.

The source current $J$ in the theory is

$$J = \tilde{f} \kappa f = \bar{f} \kappa f \ , \qquad (7)$$

where $f$ is the frame a section associated to the total field of the electron and the neutrino and $\kappa$ represents the orthonormal set. For the Sp(4,R) group the correlation reduces to conjugation.

Due to the properties that a neutral particle field should have, we consider that the effect produced by $\nu$ should be small relative to the effect of $e$ (there are no electromagnetic or massive effects). Hence we may assume that, in the composite system, $\nu$ is a perturbation of the order of the structure constant $\alpha$, the only physical constant in the theory.

$$f = e + \alpha \nu \ . \qquad (8)$$

Then the current becomes,

$$J = (\bar{e} + \alpha \bar{\nu}) \kappa (e + \alpha \nu) = \bar{e} \kappa e + \alpha (\bar{e} \kappa \nu + \bar{\nu} \kappa e) + \alpha^2 \bar{\nu} \kappa \nu \ . \qquad (9)$$

The intermediate terms may be considered as a perturbation of order $\alpha$ to a background electronic matter frame. The perturbation current may be written, by splitting $e$ into its even and odd parts and noticing that $\nu$ has only even part

$$J - \bar{e} \kappa e = \alpha \left[ \left( \bar{\eta} + \bar{\xi} \bar{\kappa}^0 \right) \kappa \nu_+ + \bar{\nu}_+ \kappa \left( \eta + \kappa^0 \xi \right) \right] \ . \qquad (10)$$

As usual in particle theory, we neglect gravitation, which is an even SL(2,C) connection. If we look for effects not imputable to gravitation, it is logical to center our attention on the odd part of the perturbation current as a candidate for the interaction current

$$\alpha j^\mu = \alpha \left( \bar{\eta} \kappa^\mu \nu + \bar{\nu} \kappa^\mu \eta \right) \ . \qquad (11)$$

It should be noted that the neutrino $\nu$ automatically associates itself, by Clifford addition, with the even part $\eta$ of the electron. This corresponds to the Weinberg Salam association of the left handed components as a doublet, with the same Lorentz transformation properties.

If we apply perturbation theory to the field equations, we expand the connection $\Gamma$ in terms of the coupling constant $\alpha$. We have,

$$J = J_0 + \alpha J_1 + \alpha^2 J_2 + \cdots , \qquad (12)$$

$$\Gamma = \Gamma_0 + \alpha \Gamma_1 + \alpha^2 \Gamma_2 + \cdots . \qquad (13)$$

The background equation and the first varied equation, which is second order in $\alpha$, have the following structure,

$$(D^*\Omega)_E = 4\pi\alpha J_E , \qquad (14)$$

$$\delta(D^*\Omega) = 4\pi\alpha \, \delta J . \qquad (15)$$

If we let, the first order terms be,

$$J_1 = j , \qquad (16)$$

$$\Gamma_1 = W , \qquad (17)$$

we obtain for the variation, the linear equation,

$$\alpha(d^*dW + LW) = 4\pi\alpha^2 j \qquad (18)$$

where $L$ is a linear first order differential operator determined by the background. This equation may be solved in principle using its Green's function $\mathcal{G}$. The solution in terms of components with respect to a base $E^a$ in the algebra is

$$W_{\mu i} = 4\pi\alpha \int dx' \mathcal{G}^{j}_{i\mu\nu}(x-x') j^{\nu}_{j}(x') . \qquad (19)$$

It is well known that the second variation of a Lagrangian serves as Lagrangian for the first varied Euler equations. Therefore, the $\Gamma J$ term in the covariant derivative present in the Lagrangian proposed in [1] provides an interaction coupling term that may be taken as part of the Lagrangian for the process in discussion. When the $\Gamma J$ term is taken in energy units, considering that the Lagrangian has an overall multiplier, it should lead to the interaction energy for the process. The second variation (or differential) in a taylor expansion of the energy U, corresponds to the Hessian of U.

$$U(0, \delta x) = U(0) + \frac{\partial U}{\partial x^i} \delta x^i + \frac{1}{2} \frac{\partial^2 U}{\partial x^j \partial x^i} \delta x^i \delta x^j + \cdots \qquad (20)$$

The $\Gamma J$ term gives the interaction Lagrangian for the background and for the perturbations, respectively,

$$\mathcal{L}_E = -\tfrac{1}{4} \mathrm{tr}\left(\tfrac{1}{2}\left[J^{\mu}_E \Gamma_{E\mu} + \Gamma_{E\mu} J^{\mu}_E\right]\right) \approx -j^{\mu}_E A_{\mu} , \qquad (21)$$

$$\mathcal{L} = -\alpha^2 \tfrac{1}{4} \mathrm{tr} \tfrac{1}{2}\left(W_{\mu} j^{\mu} + j^{\mu} W_{\mu}\right) . \qquad (22)$$

It is clear that this interaction is carried by $\Gamma$ or $W$. Nevertheless, we wish to obtain a current-current interaction to compare with other theories at low energies. Substitution of eq. (19) in the last equation gives the corresponding action which, for clarity, we indicate by,

$$\mathcal{W} = -2\pi\alpha^3 \tfrac{1}{4} \mathrm{tr} \int dx\, dx'\, \mathcal{G}^{j}_{i\mu\nu}(x-x')\left[E^i E^l + E^l E^i\right] j^{\mu}_{j}(x) j^{\nu}_{l}(x') , \qquad (23)$$

representing a current-current interaction Hamiltonian, with a coupling constant derived from the fine structure constant. This new expression may be interpreted as a weak interaction Fermi Lagrangian. The associated coupling constant is of the same order as the standard weak interaction coupling constant, up to terms in the Green's functions $\mathscr{G}$.

## Relation with Fermi's Theory

The action, in terms of elements of the algebra and the trace, corresponds to the scalar product. For the case of the isotropic homogeneous constant solution [12] the Greens function is a multiple of the unit matrix with respect to the algebra components. We assume here that for a class of solutions the Green's function have this property. Then for this class of solutions, the last equation may be written as

$$\mathscr{W} = -4\pi\alpha^3 \tfrac{1}{4} \operatorname{tr} \int dx\, dx' \, \mathscr{G}_{\mu\nu}(x-x') j^\mu(x) j^\nu(x') \quad , \tag{24}$$

where the $j$ are matrices.

The current defines an equivalent even current $j_-$ the $\kappa^0$ component, by inserting $\kappa^0\kappa^0$ and commuting, as follows,

$$j^\mu = \left(\overline{\eta}\kappa^0\overline{\kappa}^0\kappa^\mu \nu + \overline{\nu}\kappa^0\overline{\kappa}^0\kappa^\mu \eta\right) = \kappa^0\left(\eta^\dagger \overline{\sigma}^\mu \nu + \nu^\dagger \overline{\sigma}^\mu \eta\right) \quad , \tag{25}$$

$$j_- = \eta^\dagger \overline{\sigma} \nu + \text{h.c.} \quad . \tag{26}$$

Each 2×2 block component of this 4 dimensional real current matrix is an even matrix which may be represented as a complex number, by the known isomorphism between the complex algebra and a subalgebra of 2×2 matrices. Thus, we may use, instead of the generalized spinors forming the frame $f$, the standard spinor pairs, $\eta_1, \eta_2$ and $\nu_1, \nu_2$ which form the matrices corresponding to the even part, respectively, of the electron and neutrino. The first term of $j_-$, which corresponds to a non hermitian term, is

$$\eta^{\dagger \hat{B}} \overline{\sigma}^\mu \nu_A = \begin{bmatrix} \eta^{\dagger 1}\overline{\sigma}^\mu \nu_1 & \eta^{\dagger 2}\overline{\sigma}^\mu \nu_1 \\ \eta^{\dagger 1}\overline{\sigma}^\mu \nu_2 & \eta^{\dagger 2}\overline{\sigma}^\mu \nu_2 \end{bmatrix} . \tag{27}$$

In this manner, if we use the standard quantum mechanics notation, using Weyl's representation, each component is of the form

$$j^\mu = \tfrac{1}{2}\overline{\Psi}_e \gamma^\mu (1+\gamma^5)\Psi_n = \eta^\dagger \overline{\sigma}^\mu \nu \quad , \tag{28}$$

which may be recognized as the standard non hermitian Fermi weak interaction current for the electron neutrino system. We shall indicate by $j_F$ this current plus its hermitian conjugate.

As discussed in the previous publications, within this theory particles would be represented by excitations of frames. These fluctuations are matrices that correspond to a single representation of the subgroup in question. This means that for each pair of spinors of a spinor frame, only one is active for a particular fluctuation matrix. In other words, for a fluctuation, only one of the components of the matrix in eq. (27) is non zero and we may omit the indices on the spinors.

Now we may evaluate the trace of the currents in the expression for $\mathscr{W}$,

$$\operatorname{tr}(jj) = \operatorname{tr}(\kappa^0 j_- \kappa^0 j_-) = \operatorname{tr}(\overline{j}_-^\dagger \kappa^0 \kappa^0 j_-) = \operatorname{tr}(j_-^\dagger j_-) . \tag{29}$$

We see from the equations that there are various terms contributing to $\mathscr{W}$. In particular there is one term of the form,

$$\mathrm{tr}\left(\begin{bmatrix} \eta^\dagger(x)\overline{\sigma}^\mu v(x)+\mathrm{h.c.} & 0 \\ 0 & 0 \end{bmatrix}\begin{bmatrix} \eta^\dagger(x')\overline{\sigma}^\nu v(x')+\mathrm{h.c.} & 0 \\ 0 & 0 \end{bmatrix}\right) = j_0^\mu j_0^{\nu\dagger} = \\ \left(\eta^\dagger(x)\overline{\sigma}^\mu v(x)+\mathrm{h.c.}\right)\times\left(\eta^\dagger(x')\overline{\sigma}^\nu v(x')+\mathrm{h.c.}\right) \quad , \tag{30}$$

where the right side is in complex number notation instead of 2×2 even matrices. This term combines with its hermitian conjugate. It should be noticed that in going from the real 4×4 to the 2×2 complex matrix realizations the 1/4 in front of the trace changes to 1/2.

We have the result that the expression for $\mathscr{W}$ has a term of the form,

$$\mathscr{W} = -2\pi\alpha^3 \int dx\, dx'\, \mathscr{G}_{\mu\nu}(x-x') j_0^{\dagger\mu}(x) j_0^\nu(x') \quad . \tag{31}$$

If we further assume simplifications in the Green's function, which in particular are met by the fluctuations around a constant solution,

$$^*d\,^*d\delta\omega_{\hat{\delta}}^\alpha + (2m_g)^2 \delta\omega_{\hat{\delta}}^\alpha + 2(2m_g)^2 \delta\omega_{\hat{\rho}}^{\hat{\rho}}\delta_{\hat{\delta}}^\alpha = 4\pi\alpha\,\delta J_{\hat{\delta}}^\alpha \quad , \tag{32}$$

which is a Yukawa equation because of the constant background $\omega$. In general, we may expect that the Green's function has a Dirac time $\delta$ function that allows integration in $t'$. The spatial part of the Green's function should provide an equivalent range for the interaction. For example, if we assume a constant solution the equation reduces, for a point source, to the radial equation, and the Green's function is

$$\mathscr{G} = \frac{-1}{4\pi}\frac{e^{-\mu(r-r')}}{r-r'} \tag{33}$$

If we further assume that the currents vary slightly in the small region of integration so that $j(x')$ approximately equals $j(x)$, we obtain an approximation for the Hamiltonian contained in the geometric theory.

$$\mathscr{W}_w = -\frac{1}{2}\alpha^3 \int dx\, j_0^\dagger(x)\bullet j_0(x) \int_0^\infty dr'\, r\, e^{-\mu r'} \int_0^{4\pi} d^2\Omega = \\ \frac{-2\pi\alpha^3}{\mu^2}\int dx\, j_0^\dagger(x)\bullet j_0(x) \quad . \tag{34}$$

This expression fixes the value of a weak interaction constant $G$ in terms of the fine structure constant $\alpha$ and a characteristic length of the Green's function $\mathscr{G}$. The value of $G$, determined from results of experiments like the muon decay is $G=1.16639\times10^{-5}$ GeV$^{-2}$. Comparing the theoretical value with the experimental value, we may assign a mass to the value of $\omega$,

$$\frac{-2\pi\alpha^3}{\left(\mu/\cos\varphi\right)^2} = \frac{-G}{\sqrt{2}} \quad . \tag{35}$$

$$\mu = (2m_g) = 2^{3/4}\pi^{1/2}\alpha^{3/2}G^{-1/2} = 544.092 \text{ Mev.} \quad , \tag{36}$$

which corresponds to a mass close to the $\eta$ particle.

From results of previous articles, in particular for the homogeneous isotropic constant solution, using the same notation, the value of the range $R_w^{\,2}$ may be obtained theoretically in terms of a geometric fundamental unit of length, designated as g.

$$R_w^2 = 1/4\omega^2 = 1/16m^2 = 5.07694011^2/16 = 1.61095755 \text{ g}^2 \; , \quad (37)$$

The value of the geometric fundamental unit g would then be calibrated to

$$1 \text{ g}^{-1} = 0.690581 \text{ Gev} \quad (38)$$

$$1 \text{ g} = 2.85743 \times 10^{-14} \text{ cm} = 0.285743 \text{ f} \quad (39)$$

We recognize that the general perturbation for the interaction of the electron and the neutrino fields includes eqs. (28, 34) which essentially are the current and Lagrangian assumed in Fermi's theory [13, 14] of weak interactions [15, 16] of leptons. Fermi's theory is contained, as a limit within the unified theory of connections and frames discussed in these series of papers. On the other hand, the full theory is accepted there are certainly new effects and implications to consider, which should be determined without making the phenomenological considerations made in this section to display the relation of our theory to low energy weak interactions.

In particular, we should not expect that electroweak theory is related to another coupling "constant", Because of the non linearity of the theory, it is not correct to assume that if we subtract from a full solution, a partial electromagnetic solution, we get another solution. The same thing applies to strong nuclear forces. Strong and nuclear interactions were historically introduced to account for physical phenomena not explained by electromagnetic and gravitational fields. We may say that nuclear effects are residual, in the sense that they theoretically correspond to the residue of subtracting a solution to linear equation from a non linear equation

## Conclusion.

It was shown that the general perturbation technique for the interaction of the electron and neutrino fields lead to eqs. (28,34) which are essentially the current and Lagrangian assumed in Fermi's theory of weak interactions of leptons. Fermi's theory is contained, as a low energy limit within the geometric unified theory of connections and frames discussed in these series of papers. It is also clear that if the full theory is accepted there are certainly new effects.

The geometric fundamental unit of length was calibrated in terms of the accepted value for weak interaction coupling constant calculated from muon decay. From this knowledge, a value for the range of the *W* was estimated which although far from other theoretical estimates may be considered within the experimental "ballpark", without present day theoretical assumptions. Of course, we must consider in the future high energy applications which may shed some light into the relation of this geometry with the standard model [17,18].

1 G. González-Martín, Gen. Rel. and Grav. 22, 481 (1990); G. González-Martín, Physical Geometry, (Universidad Simón Bolívar, Caracas) (2000), Electronic copy posted at http:\\prof.usb.ve\ggonzalm\invstg\book.htm
2 G. González-Martín, Gen. Rel. and Grav. 23, 827 (1991).
3 G. González-Martín, Charge to Magnetic Flux Ratios in the FQHE, Universidad Simón Bolívar Report, SB/F/273-99 (1999).
4 G. González-Martín, Gen. Rel. Grav. 24, 501 (1992); See related publications.
5 E. Mach, The Science of Mechanics, 5th English ed. (Open Court, LaSalle), ch. 1 (1947).
6 A. Einstein The Meaning of Relativity, 5th ed. (Princeton Univ. Press, Princeton), p.55 (1956).
7 G. González-Martín, Fundamental Lengths in a Geometric Unified Theory, preprint 96a (1997).
8 G. González-Martín, p/e Geometric Mass Ratio, Universidad Simón Bolívar Report, SB/F/274-99 (1999).
9 G. González-Martín, The Importance of Symmetric Spaces for the Geometric Classification of Particles and Interactions, to be published as Universidad Simón Bolívar Report SB/F/279-00 (2000).
10 W. T. Grandy, Found. of Phys. 23, 439 (1993).
11 G. González-Martín, Phys. Rev. D 35, 1225 (1987).

12G. González-Martín, Fundamental Lengths in a Geometric Unified Theory, USB preprint 96a (1996).
13E, Fermi, Z. Physik 88, 161 (1934).
14E. Fermi, N. Cimento, 11 1 (1934).
15R. Feyman and M. Gell-Mann, Phys. Rev, 109, 193 (1958).
16E. C. G. Sudarshan and R. E. Marshak, Phys. Rev. 109, 1860 (1958).
17S. Weinberg, Phys. Rev. Lett. 19, 1264 (1967).
18A. Salam, in Elementary Particle Theory, ed.N. Swartholm (Almquist and Wissell, Stockholm) (1968).